# Throughput Maximization for Instantly Decodable Network Coded NOMA in Broadcast Communication Systems

Zhonghui  Mei

**Abstract :**  Non-orthogonal multiple access (NOMA) is a promising transmission scheme employed at the physical layer to improve the spectral efficiency. In this paper, we develop a novel cross-layer approach by employing NOMA at the physical layer and instantly decodable network coding (IDNC) at the network layer in downlink cellular networks. Following this approach, two IDNC packets are selected for each transmission, with one designed for all receivers and the other designed only for the strong receivers which can employ successive interference cancellation (SIC). The IDNC packets selection, transmission rates adaption for the two IDNC packets, and NOMA power allocation are jointly considered to improve the throughput of the network. Given the intractability of the problem, we decouple it into two separate subproblems, the IDNC scheduling which jointly selects the IDNC packets and the transmission rates with the given NOMA power allocation, and the NOMA power allocation with the given IDNC scheduling. The IDNC scheduling can be reduced to a maximum weight clique problem, and two heuristic algorithms named as maximum weight vertex (MWV) search and maximum weight path based maximum weight vertex (MWP-MWV) search are developed to solve the first subproblem. An iterative function evaluation (IFE) approach is proposed to solve the second subproblem. Simulation results are presented to demonstrates the throughput gain of the proposed approach over the existing solutions.

1. Introduction

Instantly decodable network coding (IDNC) is an important subclass of opportunistic network coding [1]-[2], in which the transmitter can XOR a subset or all the source packets according to the reception status and physical channel conditions fed back by the receivers, so as to enable some or all receivers to implement instant decoding. Compared with random linear network coding (RLNC) [3]-[4], IDNC has the advantages of low decoding delay, simple encoding and decoding, no requirement of buffering coded packets.

Most of the researches on IDNC focus on an upper-layer view of the network and abstract its physical-layer conditions, e.g., fading, shadowing, etc., into simple erasure channel. Moreover, they assume that all the receivers have the same physical channel capacities [5]-[11]. In [5]-[7], the authors give more priorities to the receivers with more packet requirements and higher packet loss probabilities when selecting the IDNC packets, so as to improve the throughput performance of IDNC. In [8]-[11], the authors consider selecting the IDNC packets to reduce the decoding delay, and the selected IDNC packets preferentially serve the receivers with lower packet loss probabilities.

Based on the assumption that the receivers have the same physical link capacities, the IDNC packets are selected to serve as many receivers as possible to improve throughput or decoding delay performance of the system [5]-[11]. However, due to the differences in the actual physical channel conditions of the receivers, their physical link capacities may also be different. The rate of broadcasting an IDNC packet is determined by the minimum physical link capacity among the receivers it serves. Therefore, the more receivers the IDNC packet serves, the lower the corresponding network coding broadcasting rate may be. On the other hand, the assumption that the erasure probabilities of all receivers are known at transmitter does not always hold, especially for wireless transmission channels in practical scenarios. Rate-aware instantly decodable network coding (R-IDNC) incorporates both IDNC decision at the network layer and transmission rate adaption the physical layer to improve the system performance [12]-[14]. Based on the cloud radio access networks (C-RANs), [15]-[16] jointly considers user scheduling, IDNC packet selection, rate adaption, and power allocation to maximize the system throughput. Therefore, combining IDNC technology at the network layer with wireless communication technology at the physical layer is a necessary step to make the IDNC technology from theory to practice, and can further improve the performance of IDNC.

Non-orthogonal multiple access (NOMA) has been regarded as one of the promising wireless communication technologies applied at the physical layer to increase spectrum efficiency. Compared with the conventional orthogonal multiple access (OMA) technology, the key distinguishing feature of NOMA is to support more receivers with the aid of non-orthogonal resource allocation and mitigate the effect of inter-user interference based on successive interference cancellation (SIC) [17]-[19]. Because network coding technology employed at the network layer is also an effective approach to serve multiple receivers in the same resource, a hybrid delivery scheme which includes NOMA and network coded multicasting for two-user pairing is investigated in [20]. The results in [20] reveals that network coded multicasting is the preferred choice over NOMA when the paired users have similar channel channels, and NOMA outperforms network coded multicasting when the channel gains of the pairing users are highly distinctive. However, this approach considers network coded multicasting and NOMA separately and just choose one of the two techniques by comparing their outage performance. Furthermore, the number of receivers that network coded multicasting serves is limited to be two in this paper. NOMA and RLNC are jointly considered in [21] to improve the total packet success probability in downlink cellular networks. To the best of our knowledge, in this paper, we first propose to jointly consider IDNC at the network layer and NOMA at the physical layer, so as to improve the throughput in downlink cellular networks. The main contributions of this paper are summarized as follows:

(1). We design the cross-layer optimization framework which takes the following aspects into consideration.

- IDNC packets selection: Due to employing NOMA at the physical layer, two IDNC packets are selected at the network layer during each transmission. One IDNC packet is broadcasted to all receivers, and the other IDNC packet can only be obtained by the strong receivers through employing SIC technology at the physical layer.
- Rate adaption: At the physical layer, the transmitted signals corresponding to the two IDNC packets are superimposed by employing NOMA technology. We consider an adaptive transmission rate mechanism for transmitting signals identified by each IDNC packet.
- Optimal NOMA power allocation: We also consider an optimal NOMA power allocation to maximize the throughput in the downlink cellular networks.

(2). We propose using the alternating optimization (AO) technique to solve such a difficult mixed combinatorial non-convex optimization problem. In particular, the considered optimization problem is decoupled into two subproblems named as network scheduling and power control, and the two subproblems are solved separately and iteratively.

- IDNC scheduling subproblem: Given the NOMA power allocation, the considered problem can be reduced to the network scheduling problem which includes the IDNC packets selection and the rates adaption. We first consider determining the IDNC packet and transmission rate designed for all receivers, and update the reception status according to this choice. Based on the updated reception status, we then can determine the IDNC packet and the transmission rate designed for the strong receivers. By constructing the IDNC graph, the network scheduling subproblem can be reduced to the maximum weight clique problem which is NP hard. Two heuristic algorithms named as maximum weight vertex (MWV) search and maximum weight path based maximum weight vertex (MWP-MWV) are proposed to solve the maximum weight clique problem.
- Power control subproblem: Given the network scheduling, the throughput maximization problem can be reduced to the power control subproblem which is non-convex. An iterative function evaluation (IFE) is proposed to solve this subproblem.

The remainder of this paper is organized follows. Section II introduces the system model. The throughput maximization problem for the NOMA-IDNC scheme is formulated in Section III. Section IV provides algorithms for solving the optimization problem. Section V evaluates the complexity of the proposed algorithms. Section VI and Section VII present the simulation results and conclusions, respectively.

## 2. System model

We consider a downlink cellular network where the base station (BS) broadcasts $L$ packets $\mathcal{L} = \{f_1, f_2, \cdots, f_L\}$ to $M$ receivers $\mathcal{M} = \{U_1, U_2, \cdots, U_M\}$. The BS is placed in the center of the cell, and the receivers are randomly located inside it with uniform distribution. The power level of the BS is subject to the maximum available transmit power $P_{\max}$. To enhance the system throughput, we consider adopting IDNC at the network layer, so that one IDNC packet can simultaneously serve multiple receivers. Moreover, NOMA is employed at the physical layer, so that multiple IDNC packets can be transmitted at the same time to further improve the network throughput. For simplicity, we assume a multiplexing order of 2 where two IDNC packets $Q^{(N)}$ and $Q^{(F)}$ are broadcasted for each transmission. Correspondingly, the receivers are divided into two groups: the first group $\mathcal{M}^{(N)}$ consists of the strong receivers near the BS, which can first recover $Q^{(F)}$, and then recover $Q^{(N)}$ by employing SIC; the second group $\mathcal{M}^{(F)}$ consists of the weaker receivers far away from the BS and can just receiver $Q^{(F)}$. That is to say, $Q^{(N)}$ is just designed for the strong receiver $\mathcal{M}^{(N)}$, while $Q^{(F)}$ is designed for all receivers $\mathcal{M} = \mathcal{M}^{(N)} \bigcup \mathcal{M}^{(F)}$.

**(1) NOMA**

By employing NOMA, the BS broadcasts the superposed signals of $Q^{(N)}$ and $Q^{(F)}$ at each resource in the power domain. The received signal of each receiver $U_m \in \mathcal{M}$ can be expressed as

$$y_m = h_m \left( \sqrt{P^{(N)}} x^{(N)} + \sqrt{P^{(F)}} x^{(F)} \right) + n_m \tag{1}$$

where $h_m$ is the complex channel gain from the BS to $U_m$ and is assumed to keep constant during the transmission of each superposed packet. $x^{(N)}$ and $x^{(F)}$ are the unit-power complex valued information symbol identified by the IDNC packets $Q^{(N)}$ and $Q^{(F)}$, respectively. $P^{(N)}$ indicates the transmit power of $x^{(N)}$ and $P^{(F)}$ indicates the transmit power of $x^{(F)}$. $n_m$ denotes the additive white Gaussian noise (AWGN) with variance $\sigma_m^2$.

The achievable capacity for each receiver $U_m \in \mathcal{M}$ to receive $x^{(F)}$ can be expressed as

$$R_m^{(F)} = \log_2 \left( 1 + \frac{P^{(F)} |h_m|^2}{P^{(N)} |h_m|^2 + \sigma_m^2} \right) \tag{2}$$

where $B$ is the bandwidth.

For the strong receiver $U_m \in \mathcal{M}^{(N)}$, SIC is adopted to recover $x^{(N)}$ and the corresponding achievable capacity can be represented as

$$R_m^{(N)} = \log_2 \left( 1 + \frac{P^{(N)} |h_m|^2}{\sigma_m^2} \right) \tag{3}$$

**(2) IDNC**

Similar to the past researches on IDNC [1], [2]-[14], we assume that each receiver has already received parts of source packets and buffered them. The reception status of $U_m \in \mathcal{M}$ can be indicated as follows.
- The Has set $\mathcal{H}_m \subseteq \mathcal{L}$: The set of source packet received by $U_m$.
- The Wants set $\mathcal{W}_m = \mathcal{L} \setminus \mathcal{H}_m$: The set of source packets requested by $U_m$.

We also assume that by using different channel coding rates and so on, the BS can adaptively select the transmission rates $r^{(N)}$ and $r^{(F)}$ for the signals $x^{(N)}$ and $x^{(F)}$, respectively.

Let $\mathcal{P}(Q^{(F)})$ denote the set of source packets involved in $Q^{(F)}$. $Q^{(F)}$ can be instantly decoded to retrieve a new wanted packet by $U_m \in \mathcal{M}$ if and only if the following conditions are satisfied.

(a.1) $r^{(F)} \leq R_m^{(F)}$: $U_m$ can properly receive $x^{(F)}$ if the transmission rate $r^{(F)}$ is below its achievable capacity $R_m^{(F)}$.

(a.2) $|\mathcal{P}(Q^{(F)}) \cap \mathcal{W}_m| = 1$: $U_m$ can re-XOR $Q^{(F)}$ with its previously received packets to retrieve a new packet.

Let $\tau(Q^{(F)})$ denote the set of targeted receivers of $Q^{(F)}$, that is, the set of receivers which satisfy both of the conditions (a.1) and (a.2). By exploiting $Q^{(F)}$, the reception status of $U_m$ can be updated as

$$\hat{\mathcal{W}}_m = \begin{cases} \mathcal{W}_m \setminus (\mathcal{P}(Q^{(F)}) \cap \mathcal{W}_m), & \text{if } U_m \in \tau(Q^{(F)}) \\ \mathcal{W}_m, & \text{others} \end{cases} \quad (4)$$

Similarly, the condition under which the strong receiver $U_m \in \mathcal{M}^{(N)}$ can complete instant decoding with $Q^{(N)}$, that is, $U_m \in \tau(Q^{(N)})$, can be expressed as

(b.1) $r^{(N)} \leq R_m^{(N)}$: $U_m$ can receive $x^{(N)}$ by employing SIC if the transmission rate $r^{(N)}$ is below the achievable capacity $R_m^{(N)}$.

(b.2) $|\mathcal{P}(Q^{(N)}) \cap \hat{\mathcal{W}}_m| = 1$: $Q^{(N)}$ can be instantly decoded by $U_m$ if $\mathcal{P}(Q^{(N)})$ just contains one source packet wanted by $U_m$.

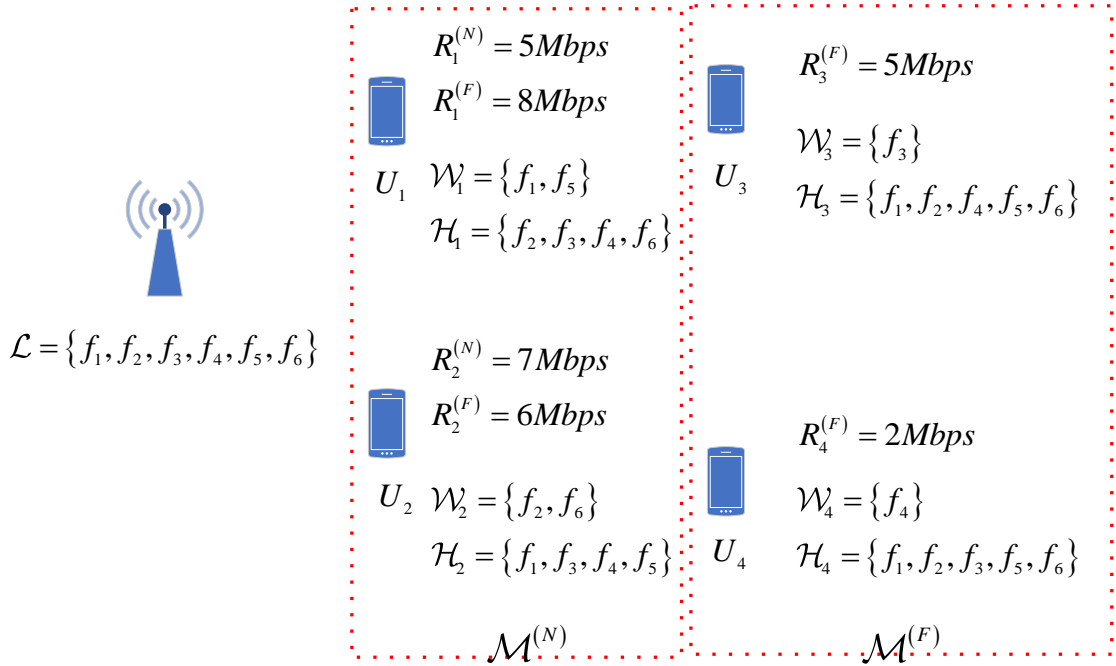

Fig. 1. An example of employing NOMA and IDNC to broadcast data packets in the downlink cellular network.

**Example 1.** This example considers a simple downlink cellular network in Fig. 1 where the BS broadcasts the source packets $\mathcal{L} = \{f_1, f_2, f_3, f_4, f_5, f_6\}$ to the receivers $\mathcal{M} = \{U_1, U_2, U_3, U_4\}$ which are divided into two groups $\mathcal{M}^{(N)} = \{U_1, U_2\}$ and $\mathcal{M}^{(F)} = \{U_3, U_4\}$. The reception status of each receiver is illustrated in Fig. 1. Assuming the selected IDNC scheduling is $Q^{(F)} = f_1 \oplus f_2 \oplus f_3$, $Q^{(N)} = P_5 \oplus P_6$, $r^{(F)} = 5bps/Hz$, and $r^{(N)} = 5bps/Hz$, by exploiting (a.1), (a.2), (b.1), and (b.2), we can get $\tau(Q^{(F)}) = \{U_1, U_2, U_3\}$ and $\tau(Q^{(N)}) = \{U_1, U_2\}$. Thus, the throughput of selected IDNC scheduling is $|\tau(Q^{(F)})| \cdot r^{(F)} + |\tau(Q^{(N)})| \cdot r^{(N)} = 25bps/Hz$. However, without employing IDNC, the achievable capacity by just employing NOMA is $15bps/Hz$ (BS concurrently transmits $x^{(F)}$ to $U_1$ with $r^{(F)} = 8bps/Hz$ and $x^{(N)}$ to $U_2$ with $r^{(N)} = 7bps/Hz$).

## 3. Problem Formulation

In this paper, the joint IDNC scheduling and NOMA power allocation to maximize the throughput of the network is carried over the variables $Q^{(F)}$, $Q^{(N)}$, $r^{(F)}$, $r^{(N)}$, $P^{(F)}$, and $P^{(N)}$, that is

$$\max_{\substack{Q^{(F)}, Q^{(N)}, r^{(F)}, r^{(N)} \\ P^{(F)}, P^{(N)}}} \left[ |\tau(Q^{(F)})| \cdot r^{(F)} + |\tau(Q^{(N)})| \cdot r^{(N)} \right] \tag{5}$$

$$s.t. \quad \tau(Q^{(F)}) = \{U_m | r^{(F)} \leq R_m^{(F)}, \\ |\mathcal{W}_m \cap \mathcal{P}(Q^{(F)})| = 1, \forall U_m \in \mathcal{M}\} \tag{5a}$$

$$\hat{\mathcal{W}}_m = \begin{cases} \mathcal{W}_m \setminus (\mathcal{P}(Q^{(F)}) \cap \mathcal{W}_m), & if \ U_m \in \tau(Q^{(F)}) \\ \mathcal{W}_m, & others \end{cases} \tag{5b}$$

$$\tau(Q^{(N)}) = \{U_m | r^{(N)} \leq R_m^{(N)}, \\ |\hat{\mathcal{W}}_m \cap \mathcal{P}(Q^{(N)})| = 1, \forall U_m \in \mathcal{M}^{(N)}\} \tag{5c}$$

$$R_m^{(F)} = \log_2 \left( 1 + \frac{P^{(F)} |h_m|^2}{P^{(N)} |h_m|^2 + \sigma_m^2} \right) \geq R_{\min}, \\ \forall U_m \in \tau(Q^{(F)}) \tag{5d}$$

$$R_m^{(N)} = \log_2 \left( 1 + \frac{P^{(N)} |h_m|^2}{\sigma_m^2} \right) \geq R_{\min}, \\ \forall U_m \in \tau(Q^{(N)}) \tag{5e}$$

$$P^{(F)} + P^{(N)} \leq P_{\max} \tag{5f}$$

where constraint (5b) represents the set of targeted receivers of $Q^{(F)}$ which meet the conditions (a.1) and (a.2). Constrain (5c) represents the updated reception status by exploiting $Q^{(F)}$.

Constraint (5d) denotes the set of targeted receivers of $Q^{(N)}$ which meet the conditions b.1 and b.2. Constraint (5e) and (5f) guarantees the minimum achievable capacity for each targeted receiver of $Q^{(F)}$ and $Q^{(N)}$. Constraint (5g) indicates the NOMA multiplexing power constrain of $P^{(F)}$ and $P^{(N)}$.

The problem in (5) is a mixed combinatorial non-convex optimization problem. The optimal solution of this problem needs to search $(2^L-1)^2 M |\mathcal{M}^{(N)}|$ combinations of IDNC packets and transmission rates, and then determine the optimal NOMA power allocation for each combination. Thus, its computational complexity increases exponentially with the number of source packets.

**Theorem 1**. NOMA-IDNC can achieve a better throughput performance than R-IDNC if the following condition satisfies

$$\left|\tau(Q^{(N)})\right| \cdot \min_{U_m \in \tau(Q^{(N)})} \log_2\left(1 + \frac{P^{(N)} |h_m|^2}{\sigma_m^2}\right)$$
$$- \left|\tau(Q^{(R\text{-}IDNC)})\right| \cdot \min_{U_i \in \tau(Q^{(R\text{-}IDNC)})} \log_2\left(1 + \frac{P^{(N)} |h_i|^2}{\sigma_i^2}\right) \quad (6)$$
$$> 0$$

where $Q^{(R-IDNC)}$ is the optimal IDNC packet selected for R-IDNC, and $\tau(Q^{(R-IDNC)})$ is the set of targeted receivers of $Q^{(R-IDNC)}$.

**Proof:** Assuming the optimal IDNC packet which can achieve the maximum throughput for R-IDNC is $Q^{(R-IDNC)}$ with $\tau(Q^{(R-IDNC)})$ and $r^{(R-IDNC)} = \min_{u_i \in \tau(Q^{(R-IDNC)})} \log_2\left(1 + \frac{P_{\max} |h_i|^2}{\sigma_i^2}\right)$, the maximum throughput achieved by R-IDNC can be expressed as

$$R^{(R\text{-}IDNC)} = \left|\tau(Q^{(R\text{-}IDNC)})\right| \cdot \min_{u_i \in \tau(Q^{(R\text{-}IDNC)})} \log_2\left(1 + \frac{P_{\max} |h_i|^2}{\sigma_i^2}\right) \quad (7)$$

In NOMA-IDNC, we may select $Q^{(F)} = Q^{(R-IDNC)}$ with $\tau(Q^{(F)}) = \tau(Q^{(R-IDNC)})$ $r^{(F)} = \min_{u_i \in \tau(Q^{(F)})} \log_2\left(1 + \frac{(P_{\max} - P^{(N)})|h_i|^2}{P^{(N)} |h_i|^2 + \sigma_i^2}\right)$, and $Q^{(N)}$ with $\tau(Q^{(N)})$ and $r^{(N)} = \min_{u_m \in \tau(Q^{(N)})} \log_2\left(1 + \frac{P^{(N)} |h_m|^2}{\sigma_m^2}\right)$.

Thus, the throughput achieved by NOMA-IDNC can be expressed as

$$R^{(NOMA-IDNC)}$$
$$= \left|\tau\left(Q^{(R-IDNC)}\right)\right|$$
$$\times \min_{u_i \in \tau\left(Q^{(R-IDNC)}\right)} \log_2\left(1 + \frac{\left(P_{max} - P^{(N)}\right)|h_i|^2}{P^{(N)}|h_i|^2 + \sigma_i^2}\right) \quad (8)$$
$$+ \left|\tau\left(Q^{(N)}\right)\right| \times \min_{u_m \in \tau\left(Q^{(N)}\right)} \log_2\left(1 + \frac{P^{(N)}|h_m|^2}{\sigma_m^2}\right)$$

It is obvious that

$$\arg \min_{u_i \in \tau\left(Q^{(R-IDNC)}\right)} \log_2\left(1 + \frac{P_{max}|h_i|^2}{\sigma_i^2}\right)$$
$$= \arg \min_{u_i \in \tau\left(Q^{(R-IDNC)}\right)} \log_2\left(1 + \frac{\left(P_{max} - P^{(N)}\right)|h_i|^2}{P^{(N)}|h_i|^2 + \sigma_i^2}\right) \quad (9)$$

By comparing (7) and (8), we can

$$R^{(NOMA-IDNC)} - R^{(R-IDNC)}$$
$$= \left|\tau\left(Q^{(N)}\right)\right| \cdot \min_{u_m \in \tau\left(Q^{(N)}\right)} \log_2\left(1 + \frac{P^{(N)}|h_m|^2}{\sigma_m^2}\right) \quad (10)$$
$$- \left|\tau\left(Q^{(R-IDNC)}\right)\right| \cdot \min_{U_i \in \tau\left(Q^{(R-IDNC)}\right)} \log_2\left(1 + \frac{P^{(N)}|h_i|^2}{\sigma_i^2}\right)$$

## 4. Solution of the Optimization Problem

In order to reach a tractable solution to (5), a convenient approach is to decouple it into the IDNC schedule subproblem and the power control subproblem, and employ the alternating optimization technique to separately and iteratively solve the two subproblems.

**(1) IDNC schedule subproblem**

Given the NOMA power allocation, the optimization problem (5) can be formulated as the IDNC schedule problem, which can be expressed as

$$\max_{Q^{(N)}, Q^{(F)}, r^{(N)}, r^{(F)}} \left[\left|\tau\left(Q^{(F)}\right)\right| \cdot r^{(F)} + \left|\tau\left(Q^{(N)}\right)\right| \cdot r^{(N)}\right] \quad (11)$$
$$s.t. \quad (5b),(5c),(5d)$$

Problem (10) is a combinatorial optimization problem. The solution of the optimal problem requires an exhaustive search for $(2^L - 1)^2 M \left|\mathcal{M}^{(N)}\right|$ combinations of IDNC packets and transmission rates. In the following, we reformulate the problem (10) by using the graph-theoretic techniques.

Firstly, we design the IDNC graph $\mathcal{G}\left(\mathcal{V}^{(F)}, \mathcal{E}^{(F)}\right)$ by exploiting the reception status and the set of achievable capacities $\mathcal{R}^{(F)} = \left\{R_m^{(F)}, U_m \in \mathcal{M}\right\}$ corresponding to $Q^{(F)}$. $\mathcal{V}^{(F)}$ and $\mathcal{E}^{(F)}$ are the set of vertices and edges of the graph, respectively.

**Definition 1.** $v_{m,l,r} \in \mathcal{V}^{(F)}$ is created for $\forall f_l \in \mathcal{W}_m$, $r \in \mathcal{R}_m^{(F)} = \left\{r \mid r \leq R_m^{(F)}, \quad r \in \mathcal{R}^{(F)}\right\}$, and $U_m \in \mathcal{M}$.

**Definition 2.** $\varepsilon(v_{m,l,r}, v_{m',l',r'}) \in \mathcal{E}^{(F)}$ is generated by connecting two vertices $v_{m,l,r}$ and $v_{m',l',r'}$ if they satisfy both of the following conditions:

(c.1) $r = r'$, which ensures the transmission rate is the same for all adjacent vertices in the IDNC graph.

(c.2) $(l = l', m \neq m')m \neq m')$ or $(((f_l, f_{l'}) \in \mathcal{H}_{m'} \times \mathcal{H}_m, m \neq m'), m \neq m')$, which represents the instant decodable condition of IDNC from the network layer perspective.

Secondly, by exploiting the updated reception status $\{\hat{\mathcal{W}}_m, U_m \in \mathcal{M}^{(N)}\}$ and the achievable capacities $\mathcal{R}^{(N)} = \{R_m^{(N)}, U_m \in \mathcal{M}^{(N)}\}$, we can generate the IDNC graph $\mathcal{G}(\mathcal{V}^{(N)}, \mathcal{E}^{(N)})$ where the vertices and edges can be defined as follows.

**Definition 3.** $v_{m,l,r} \in \mathcal{V}^{(N)}$ is produced for $\forall P_l \in \hat{\mathcal{W}}_m$, $r \in \mathcal{R}_m^{(N)} = \{r | r \leq R_m^{(N)}, r \in \mathcal{R}^{(N)}\}$, and $U_m \in \mathcal{M}^{(N)}$.

**Definition 4.** An edge $\varepsilon(v_{m,l,r}, v_{m',l',r'}) \in \mathcal{E}^{(N)}$ is created if both of the conditions (c.1) and (c.2) are satisfied.

**Example 2.** The IDNC graph $\mathcal{G}(\mathcal{V}^{(F)}, \mathcal{E}^{(F)})$ corresponding to Example 1 is shown in Fig. 2. Assuming that $Q^{(F)} = P_1 \oplus P_2 \oplus P_3$ and $r^{(F)} = 5Mbps$, the updated reception status is $\hat{\mathcal{W}}_1 = \{P_5\}$, $\hat{\mathcal{W}}_2 = \{P_6\}$, $\hat{\mathcal{W}}_3 = \varnothing$, and $\hat{\mathcal{W}}_4 = \{P_4\}$. Thus, we can get another IDNC graph $\mathcal{G}(\mathcal{V}^{(N)}, \mathcal{E}^{(N)})$ shown in Fig. 3.

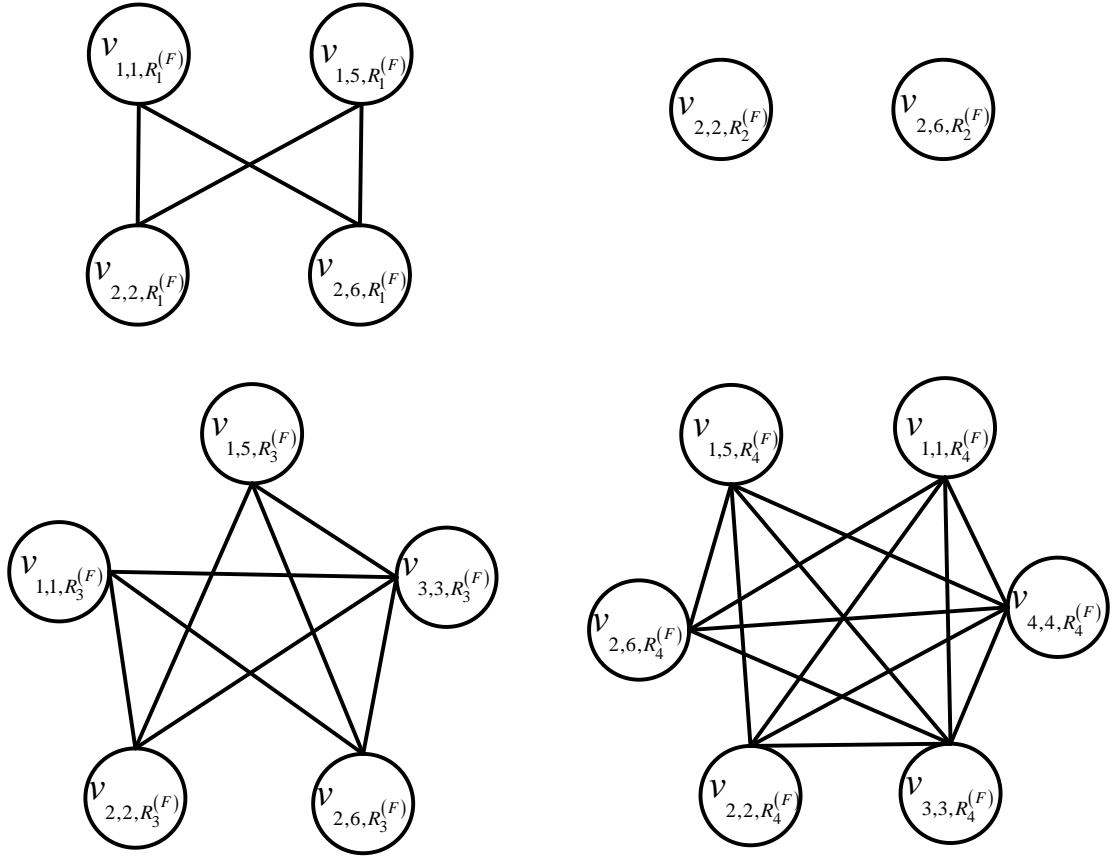

**Fig. 2.** $\mathcal{G}(\mathcal{V}^{(F)}, \mathcal{E}^{(F)})$ **of Example 1.**

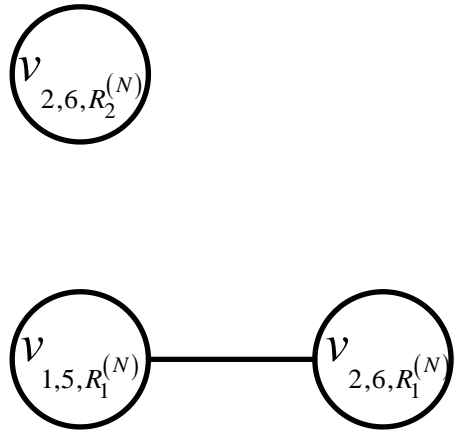

**Fig. 3.** $\mathcal{G}(\mathcal{V}^{(N)}, \mathcal{E}^{(N)})$ **of Example 1 with** $Q^{(F)} = P_1 \oplus P_2 \oplus P_3$ **and** $r^{(F)} = 5Mbps$.

**Definition 5.** A clique is a set of vertices in the IDNC graph where any two vertices can be connected by an edge.

**Definition 6.** A maximal clique is defined as a clique that is not s subset of any larger clique.

We define the weight of each vertex of the IDNC graphs $\mathcal{G}(\mathcal{V}^{(F)}, \mathcal{E}^{(F)})$ and $\mathcal{G}(\mathcal{V}^{(N)}, \mathcal{E}^{(N)})$ as

$$\omega(v_{m,l,r}) = r, \quad \forall v_{m,l,r} \in \mathcal{V}^{(F)} \cup \mathcal{V}^{(N)} \tag{12}$$

Obviously, with the above definitions, the set of feasible solutions of $\{Q^{(F)}, r^{(F)}\}$ and $\{Q^{(N)}, r^{(N)}\}$ can be characterized by the cliques in $\mathcal{G}(\mathcal{V}^{(F)}, \mathcal{E}^{(F)})$ and $\mathcal{G}(\mathcal{V}^{(N)}, \mathcal{E}^{(N)})$, respectively. The cliques which can achieve the maximum throughput lies in the maximal clique set. Assuming that the set of available maximum cliques in $\mathcal{G}(\mathcal{V}^{(F)}, \mathcal{E}^{(F)})$ is $\mathcal{K}^{(F)}$, and given the IDNC schedule $\{Q^{(F)}, r^{(F)}\}$ identified by the maximum clique $\kappa^{(F)}$, the set of available maximum cliques in $\mathcal{G}(\mathcal{V}^{(N)}, \mathcal{E}^{(N)})$ is $\mathcal{K}^{(N)}|\kappa^{(F)}$, the optimization problem (11) can be formulated to the maximum weight clique problem, that is

$$\max_{\substack{\kappa^{(F)} \in \mathcal{K}^{(F)}, \\ \kappa^{(N)} \in \mathcal{K}^{(N)}|\kappa^{(F)}}} \left[ \sum_{v_{m',l',r'} \in \kappa^{(F)}} \omega(v_{m',l',r'}) + \sum_{v_{m,l,r} \in \kappa^{(N)}} \omega(v_{m,l,r}) \right] \tag{13}$$

The above optimization problem requires searching $\sum_{\kappa^{(F)} \in \mathcal{K}^{(F)}} \left|\left(\mathcal{K}^{(N)}|\kappa^{(F)}\right)\right|$ maximum cliques. To reduce the complexity, we consider first determining the maximum weight clique $\kappa^{*(F)} = \arg\max_{\kappa^{(F)} \in \mathcal{K}^{(F)}} \sum_{v_{m,l,r} \in \kappa^{(F)}} \omega(v_{m,l,r})$, and then determining the maximum weight clique $\kappa^{*(N)}$ selected from $\mathcal{K}^{(N)}|\kappa^{*(F)}$, that is

$$\begin{aligned}
&\max_{\substack{\kappa^{(F)} \in \mathcal{K}^{(F)}, \\ \kappa^{(N)} \in \mathcal{K}^{(N)}|\kappa^{(F)}}} \left[ \sum_{v_{m',l',r'} \in \kappa^{(F)}} \omega(v_{m',l',r'}) + \sum_{v_{m,l,r} \in \kappa^{(N)}} \omega(v_{m,l,r}) \right] \\
&\approx \sum_{v_{m',l',r'} \in \kappa^{*(F)}} \omega(v_{m',l',r'}) + \max_{\kappa^{(N)} \in \mathcal{K}^{(N)}|\kappa^{*(F)}} \sum_{v_{m,l,r} \in \kappa^{(N)}} \omega(v_{m,l,r})
\end{aligned} \tag{14}$$

Thus, the above approximate optimization problem is equal to searching two separate maximum weight cliques, that is

$$\kappa^{*(F)} = \arg\max_{\kappa^{(F)} \in \mathcal{K}^{(F)}} \sum_{v_{m,l,r} \in \kappa^{(F)}} \omega(v_{m,l,r}) \tag{15}$$

$$\kappa^{*(N)} = \arg\max_{\kappa^{(N)} \in \mathcal{K}^{(N)}|\kappa^{*(F)}} \sum_{v_{m,l,r} \in \kappa^{(N)}} \omega(v_{m,l,r}) \tag{16}$$

It is well known that the maximum weight clique problem is NP hard [1], [3]-[14]. Since the methods for solving the maximum weight clique problem (14) and (15) are the same, we will only consider the solution to the maximum weight clique problem (14) below.

Let $\kappa^{*(F)}_{\{v_{m,l,r}\}}$ denote the maximum weight clique containing $v_{m,l,r}$, that is, it is the maximal clique which contains $v_{m,l,r}$ and has the maximum weight. The weight of $\kappa^{*(F)}_{\{v_{m,l,r}\}}$ can be expressed as

$$\omega\left(\kappa^{*(F)}_{\{v_{m,l,r}\}}\right) = \sum_{v_{m',l',r'} \in \kappa^{*(F)}_{\{v_{m,l,r}\}}} \omega\left(v_{m',l',r'}\right) \tag{17}$$

The optimization problem (14) can be reformulated as

$$\begin{cases} v^* = \arg\max_{v_{m,l,r} \in \mathcal{V}^{(F)}} \omega\left(\kappa^{*(F)}_{\{v_{m,l,r}\}}\right) \\ \kappa^{*(F)} = \kappa^{*(F)}_{\{v^*\}} \end{cases} \tag{18}$$

By modifying the vertex weight of $v_{m,l,r}$ to $\omega\left(\kappa^{*(F)}_{\{v_{m,l,r}\}}\right)$, the MWC problem is equal to searching the vertex with the maximum modified vertex weight. However, computing $\omega\left(\kappa^{*(F)}_{\{v_{m,l,r}\}}\right)$ involves searching $\kappa^{*(F)}_{\{v_{m,l,r}\}}$, which is also NP hard. In the following, we propose two heuristic algorithms to approximate $\omega\left(\kappa^{*(F)}_{\{v_{m,l,r}\}}\right)$.

**(a) MWV algorithm**

To the best of our knowledge, in the past research on IDNC, MWV has often been used to solve the maximum weight clique problem [1], [3]-[12]. In this algorithm, it approximates $\omega\left(\kappa^{*(F)}_{\{v_{m,l,r}\}}\right)$ by considering the following two aspects:

(d.1) When $\omega\left(v_{m,l,r}\right)$ is large, $\omega\left(\kappa^{*(F)}_{\{v_{m,l,r}\}}\right)$ may also be large.

(d.2) A larger value of $\sum_{v_{m',l',r'} \in \mathcal{C}(v_{m,l,r})} \omega\left(v_{m',l',r'}\right)$ may lead to a larger value of $\omega\left(\kappa^{*(F)}_{\{v_{m,l,r}\}}\right)$, where $\mathcal{C}\left(v_{l,m,r}\right)$ denotes the set of adjacent vertices of $v_{l,m,r}$ in $\mathcal{G}\left(\mathcal{V}^{(F)}, \mathcal{E}^{(F)}\right)$.

Thus, according to MWV, $\omega\left(\kappa^{*(F)}_{\{v_{m,l,r}\}}\right)$ can be approximated to

$$\begin{aligned} &\omega\left(\kappa^{*(F)}_{\{v_{m,l,r}\}}\right) \\ &\approx \tilde{\omega}\left(v_{m,l,r}\right) \\ &= \omega\left(v_{m,l,r}\right) \cdot \sum_{v_{m',l',r'} \in \mathcal{C}(v_{m,l,r})} \omega\left(v_{m',l',r'}\right) \end{aligned} \tag{19}$$

For convenience, we summarize MWV in Algorithm 1.

**Algorithm 1.** MWV algorithm

Initialize: $\mathcal{V}_{sel} = \mathcal{V}^{(F)}$, $\kappa^{*(F)} = \varnothing$;

While $\mathcal{V}_{sel} \neq \varnothing$

    Compute $\tilde{\omega}(v_{m,l,r})$ with (18), $\forall v_{m,l,r} \in \mathcal{V}_{sel}$;

    Select $\tilde{v}^* = \arg\max_{v_{m,l,r} \in \mathcal{V}_{sel}} \tilde{\omega}(v_{m,l,r})$;

    Update $\kappa^{*(F)} \Leftarrow \kappa^{*(F)} \bigcup \tilde{v}^*$;

    Update $\mathcal{V}_{sel} \Leftarrow \mathcal{V}_{sel} \cap \mathcal{C}(\tilde{v}^*)$;

End while

Output: $\kappa^{*(F)}$.

However, in MWV, the condition e.2 may be true only if $\mathcal{C}(v_{l,m,r})$ can form a clique, that is, any two vertices in $\mathcal{C}(v_{l,m,r})$ can be connected by an edge. Otherwise, in (18), $\tilde{\omega}(v_{m,l,r})$ cannot indicate $\omega(\kappa^{*(F)}_{\{v_{m,l,r}\}})$ very well, which may lead to the wrong selection of the maximum weight clique.

**(b) MWP-MWV algorithm**

**Definition 5.** A maximal weight path (MWP) originated from $v_{l,m,r}$ can be denoted as

$$\mathbb{L}(v_{l,m,r}) = v_{l,m,r} \to v_{l_1,m_1,r_1} \to \cdots \to v_{l_x,m_x,r_x}, \quad \text{where} \quad v_{l_k,m_k,r_k} = \arg\max_{v_{l',m',r'} \in \mathcal{C}(v_{l,m,r}) \bigcap_{n=1}^{k-1} \mathcal{C}(v_{l_n,m_n,r_n})} \omega(v_{l',m',r'}),$$

$k \in \{1,2,\cdots,x\}$, $\mathcal{C}(v_{m,l,r}) \bigcap_{n=1}^{x} \mathcal{C}(v_{m_n,l_n,r_n}) = \varnothing$.

**Lemma 1.** $\mathbb{L}(v_{l,m,r})$ is a maximal clique.

**Proof.** Without loss of generality, assuming that $\mathbb{L}(v_{l,m,r}) = v_{l,m,r} \to v_{l_1,m_1,r_1} \to \cdots \to v_{l_x,m_x,r_x}$, we can get $v_{l_k,m_k,r_k} = \arg\max_{v_{l',m',r'} \in \mathcal{C}(v_{l,m,r}) \bigcap_{n=1}^{k-1} \mathcal{C}(v_{l_n,m_n,r_n})} \omega(v_{l',m',r'})$ ($k \in \{1,2,\cdots,x\}$) according to Definition 5. Therefore, $v_{l_k,m_k,r_k}$ ($k \in \{1,2,\cdots,x\}$) has an edge with any vertex of $\{v_{l,m,r}, v_{l_1,m_1,r_1}, \cdots, v_{l_{k-1},m_{k-1},r_{k-1}}\}$, that is, $\mathbb{L}(v_{l,m,r})$ forms a clique.

On the other hand, $\mathcal{C}(v_{m,l,r}) \bigcap_{n=1}^{x} \mathcal{C}(v_{m_n,l_n,r_n}) = \varnothing$ ensures that $\mathbb{L}(v_{l,m,r})$ is a maximal clique.

**Proposition 1.** $|\mathbb{L}(v_{m,l,r})| \leq M$.

**Proof**: For $\forall v_{m',l',r'}, v_{m'',l'',r''} \in \mathbb{L}(v_{m,l,r})$, according to Lemma 1 and condition C.2, we can get $m' \neq m''$, which concludes the proof.

Therefore, in MWP-MWV, $\omega\left(\kappa_{\{v_{m,l,r}\}}^{*(F)}\right)$ can be approximated with the weight of $\mathbb{L}(v_{l,m,r})$, that is

$$\omega\left(\kappa_{\{v_{m,l,r}\}}^{*(F)}\right) \approx \hat{\omega}(v_{m,l,r}) = \sum_{v_{m',l',r'} \in \mathbb{L}(v_{m,l,r})} \omega(v_{m',l',r'}) \tag{20}$$

The MWP-MWV algorithm is summarized in Algorithm 2.

**Algorithm 2.** MWP-MWV

For $\forall v_{m,l,r} \in \mathcal{V}^{(F)}$

    Initialize $\mathbb{L}(v_{l,m,r}) = v_{l,m,r}$ and $\mathcal{V}_{sel} = \mathcal{C}(v_{m,l,r})$

    While $\mathcal{V}_{sel} \neq \varnothing$

        Select $\hat{v}^* = \arg \max_{v_{m,l,r} \in \mathcal{V}_{sel}} \omega(v_{m,l,r})$;

        Update $\mathbb{L}(v_{l,m,r}) \Leftarrow \mathbb{L}(v_{l,m,r}) \rightarrow \hat{v}^*$;

        Update $\mathcal{V}_{sel} \Leftarrow \mathcal{V}_{sel} \cap \mathcal{C}(\hat{v}^*)$;

    End while

    Compute $\hat{\omega}(v_{m,l,r}) = \sum_{v_{m',l',r'} \in \mathbb{L}(v_{l,m,r})} \omega(v_{m',l',r'})$;

End for

Output: $v^* = \arg \max_{v_{m,l,r} \in \mathcal{V}^{(F)}} \hat{\omega}(v_{m,l,r})$, $\kappa^{*(F)} = \mathbb{L}(v^*)$.

By employing MWV algorithm or MWP-MWV algorithm, the optimization problems (14) and (15) can be solved with Algorithm 3, which is summarized as follows.

**Algorithm 3**. Compute $\kappa^{*(F)}$ and $\kappa^{*(N)}$ with MWV (MWP-MWV)

Construct $\mathcal{G}\left(\mathcal{V}_m^{(F)}, \mathcal{E}_m^{(F)}\right)$;

Compute $\kappa^{*(F)}$ with MWV (MWP-MWV);

Update $\hat{\mathcal{W}}_m = \mathcal{W}_m \setminus P_l$, $\forall v_{m,l,r} \in \kappa^{*(F)}$;

Construct $\mathcal{G}\left(\mathcal{V}_m^{(N)}, \mathcal{E}_m^{(N)}\right)$ with the updated reception status;

Compute $\kappa^{*(N)}$ with MWV (MWP-MWV);

**(2) Power control subproblem**

Given the resulting IDNC schedule $\kappa^{*(F)}$ and $\kappa^{*(N)}$, the objective function of the problem (5) is optimized over the NOMA power allocation $P^{(F)}$ and $P^{(N)}$, which can be expressed as

$$\max_{P^{(F)},\, P^{(N)}} \left[ \left|\kappa^{*(F)}\right| \cdot \min_{U_m \in \tau\left(\kappa^{*(F)}\right)} R_m^{(F)} \right.$$
$$\left. + \left|\kappa^{*(N)}\right| \cdot \min_{U_m \in \tau\left(\kappa^{*(N)}\right)} R_m^{(N)} \right] \quad (21)$$
$$s.t. \quad (5e), (5f), (5g)$$

The above power control problem is a non-convex problem. Therefore, similar to [19]-[20], we focus on achieving at least a local optimum solution using the IFE approach.

According to (5e), we can get

$$P^{(F)} \geq \left( P^{(N)} + \frac{\sigma_m^2}{|h_m|^2} \right)\left( 2^{R_{\min}} - 1 \right), \quad \forall U_m \in \tau\left(\kappa^{*(F)}\right) \quad (22)$$

By defining $U_{x_f} = \arg\max_{U_m \in \tau(\kappa^{*(F)})} \frac{\sigma_m^2}{|h_m|^2}$, we can rewrite (21) as

$$P^{(F)} \geq \left( P^{(N)} + \frac{\sigma_{x_f}^2}{\left|h_{x_f}\right|^2} \right)\left( 2^{R_{\min}} - 1 \right) \quad (23)$$

Similarly, by defining $U_{x_n} = \arg\max_{U_m \in \tau(\kappa^{*(N)})} \frac{\sigma_m^2}{|h_m|^2}$, we can rewrite (5f) as

$$P^{(N)} \geq \frac{\sigma_{x_n}^2}{\left|h_{x_n}\right|^2}\left( 2^{R_{\min}} - 1 \right) \quad (24)$$

Thus, with (22), (23), and (5g), we can show the feasible solution domain of the optimization problem (20) in Fig. 4.

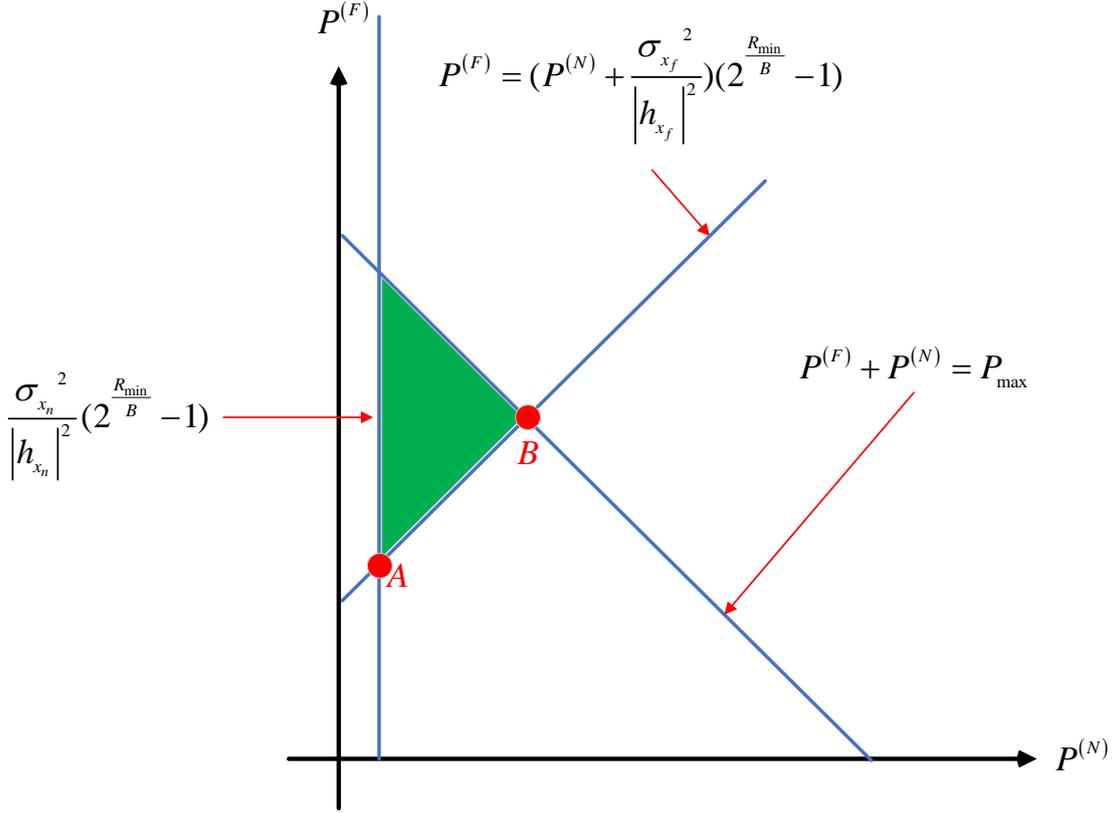

**Fig. 4.** Feasible solution domain of the optimization problem (21).

**Lemma 2.** The optimization problem (20) has feasible solutions if and only if

$$P_{\max} \geq \frac{\sigma_{x_f}^2}{|h_{x_f}|^2} \cdot \left(2^{R_{\min}} - 1\right) + \frac{\sigma_{x_n}^2}{|h_{x_n}|^2} \cdot 2^{R_{\min}} \left(2^{R_{\min}} - 1\right) \tag{25}$$

**Proof:** As is shown in Fig. 4, the coordinates of point A is $\left(\frac{\sigma_{x_n}^2}{|h_{x_n}|^2} \cdot \left(2^{R_{\min}} - 1\right), \frac{\sigma_{x_n}^2}{|h_{x_n}|^2} \cdot \left(2^{R_{\min}} - 1\right)^2 + \frac{\sigma_{x_f}^2}{|h_{x_f}|^2} \cdot \left(2^{R_{\min}} - 1\right)\right)$, and the coordinates of point B is $\left(P_{\max} \cdot 2^{-R_{\min}} - \frac{\sigma_{x_f}^2}{|h_{x_f}|^2} \cdot \left(1 - 2^{-R_{\min}}\right), P_{\max} - P_{\max} \cdot 2^{-R_{\min}} + \frac{\sigma_{x_f}^2}{|h_{x_f}|^2} \cdot \left(1 - 2^{-R_{\min}}\right)\right)$. The condition that optimization problem (20) has feasible solutions is that the horizontal coordinate value of point A is less than that of point B, that is

$$\frac{\sigma_{x_n}^2}{|h_{x_n}|^2} \cdot \left(2^{R_{\min}} - 1\right) \leq P_{\max} \cdot 2^{-R_{\min}} - \frac{\sigma_{x_f}^2}{|h_{x_f}|^2} \cdot \left(1 - 2^{-R_{\min}}\right) \tag{26}$$

Thus, with (25), we can get the conclusions of Lemma 2.

**Lemma 3.** The feasible solution domain of $P^{(N)}$ in problem (20) is $\left[\left(P^{(N)}\right)^{(\text{LOW})}, \left(P^{(N)}\right)^{(\text{UP})}\right]$, where

$$\left(P^{(N)}\right)^{(\text{LOW})} = \frac{\sigma_{x_n}^2}{|h_{x_n}|^2} \cdot \left(2^{R_{\min}} - 1\right) \text{ and } \left(P^{(N)}\right)^{(\text{UP})} = P_{\max} \cdot 2^{-R_{\min}} - \frac{\sigma_{x_f}^2}{|h_{x_f}|^2} \cdot \left(1 - 2^{-R_{\min}}\right).$$

**Proof:** According to Fig. 4, we can get the feasible solution domain of $P^{(N)}$ is between the horizontal coordinate values of point A and point B. Therefore, we can get the conclusions of Lemma 3.

According to the optimization problem (20), given $P^{(N)}$, the larger $P^{(F)}$, the larger throughput can be achieved accordingly. Therefore, the optimization problem (20) can be reduced to

$$\max_{P^{(N)}} \left[ |\kappa^{*(F)}| \cdot \log_2\left(1 + \frac{\left(P_{\max} - P^{(N)}\right)|h_{x_f}|^2}{P^{(N)}|h_{x_f}|^2 + \sigma_{x_f}^2}\right) + |\kappa^{*(N)}| \cdot \log_2\left(1 + \frac{P^{(N)}|h_{x_n}|^2}{\sigma_{x_n}^2}\right) \right] \quad (27)$$

$$s.t. \quad \left(P^{(N)}\right)^{(\text{low})} \le P^{(N)} \le \left(P^{(N)}\right)^{(\text{up})} \quad (26a)$$

**Lemma 4.** Given the IDNC scheme $\kappa^{*(F)}$ and $\kappa^{*(N)}$, a closed-form of the power $P^{(N)}$ can be obtained by updating $P^{(N)}$ at iteration $(k+1)$ according to the following equation

$$\left(P^{(N)}\right)^{(k+1)} = \left[ \frac{|\kappa^{*(N)}| \cdot \dfrac{\text{SINR}_{x_n}}{1+\text{SINR}_{x_n}}}{|\kappa^{*(F)}| \cdot \dfrac{\left(\text{SINR}_{x_f}\right)^2}{1+\text{SINR}_{x_f}} \cdot \dfrac{|h_{x_f}|^2}{P_{\max}|h_{x_f}|^2 + \sigma_{x_f}^2}} \right]_{\left(P^{(N)}\right)^{(\text{low})}}^{\left(P^{(N)}\right)^{(\text{up})}}, \quad (28)$$

where $\text{SINR}_{x_n} = \dfrac{\left(P^{(N)}\right)^{(k)}|h_{x_n}|^2}{\sigma_{x_n}^2}$; $\text{SINR}_{x_f} = \dfrac{\left(P_{\max} - \left(P^{(N)}\right)^{(k)}\right)|h_{x_f}|^2}{\left(P^{(N)}\right)^{(k)}|h_{x_f}|^2 + \sigma_{x_f}^2}$; $\left(P^{(N)}\right)^{(k)}$ is the transmit power at the $k$-th iteration.

**Proof.** According the objective function (26), we can define

$$\phi\left(P^{(N)}\right) = |\kappa^{*(F)}| \cdot \log_2\left(1 + \frac{\left(P_{\max} - P^{(N)}\right)|h_{x_f}|^2}{P^{(N)}|h_{x_f}|^2 + \sigma_{x_f}^2}\right) + |\kappa^{*(N)}| \cdot \log_2\left(1 + \frac{P^{(N)}|h_{x_n}|^2}{\sigma_{x_n}^2}\right) \quad (29)$$

By taking the first derivative of $\phi(P^{(N)})$ with respective to $P^{(N)}$, we can obtain

$$\begin{aligned}
&\frac{\partial}{\partial P^{(N)}}\phi(P^{(N)}) \\
&= \frac{\partial}{\partial P^{(N)}}\left[\left|\kappa^{*(F)}\right|\cdot\log_2\left(1+\frac{(P_{\max}-P^{(N)})\left|h_{x_f}\right|^2}{P^{(N)}\left|h_{x_f}\right|^2+\sigma_{x_f}^2}\right)\right. \\
&\quad \left.+\left|\kappa^{*(N)}\right|\cdot\log_2\left(1+\frac{P^{(N)}\left|h_{x_n}\right|^2}{\sigma_{x_n}^2}\right)\right] \\
&= \frac{\partial}{\partial P^{(N)}}\left[\left|\kappa^{*(F)}\right|\cdot\log_2\left(1+\mathrm{SINR}_{x_f}\right)\right. \\
&\quad \left.+\left|\kappa^{*(N)}\right|\cdot\log_2\left(1+\mathrm{SINR}_{x_n}\right)\right] \\
&= -\left|\kappa^{*(F)}\right|\cdot\frac{\left(\mathrm{SINR}_{x_f}\right)^2}{1+\mathrm{SINR}_{x_f}}\cdot\frac{\left|h_{x_f}\right|^2}{P_{\max}\left|h_{x_f}\right|^2+\sigma_{x_f}^2} \\
&\quad +\left|\kappa^{*(N)}\right|\cdot\frac{\mathrm{SINR}_{x_n}}{1+\mathrm{SINR}_{x_n}}\cdot\frac{1}{P^{(N)}}
\end{aligned} \qquad (30)$$

By letting $\frac{\partial}{\partial P^{(N)}}\phi(P^{(N)})=0$, we can get

$$P^{(N)} = \frac{\left|\kappa^{*(N)}\right|\cdot\frac{\mathrm{SINR}_{x_n}}{1+\mathrm{SINR}_{x_n}}}{\left|\kappa^{*(F)}\right|\cdot\frac{\left(\mathrm{SINR}_{x_f}\right)^2}{1+\mathrm{SINR}_{x_f}}\cdot\frac{\left|h_{x_f}\right|^2}{P_{\max}\left|h_{x_f}\right|^2+\sigma_{x_f}^2}} \qquad (31)$$

Finally, by taking account of the constraint (26b), an additional projection step on the constraint set $[\,\cdot\,]_{(P^{(N)})^{(\mathrm{low})}}^{(P^{(N)})^{(\mathrm{up})}}$ must be included in each iteration, which completes the proof of Lemma 4.

It must be noted that the iterative process of (3) converges to a local stationary point which falls inside of the constrained set or on the boundary, so as to satisfy the KKT conditions of optimality [19]-[20].

Therefore, the overall two-step algorithm to solve (5) can be summarized in Algorithm 4.

**Algorithm 4.** NOMA-IDNC，MWV （NOMA-IDNC, MWP-MWV）
**While** problem (5) has not converged **do**
    Compute $\kappa^{*(F)}$ and $\kappa^{*(N)}$ based on algorithm 3 with **MWV (MWP-MWV)**.
    **While** problem (26) has not converged **do**
     Update $P^{(N)}$ with (27).
    **End while**
**End while**

## 5. Complexity Analysis

In this section, we evaluate the complexity of our solution to IDNC schedule presented in Algorithm 3. The complexity of this algorithm consists of constructing the IDNC graphs $\mathcal{G}(\mathcal{V}^{(F)}, \mathcal{E}^{(F)})$ and $\mathcal{G}(\mathcal{V}^{(N)}, \mathcal{E}^{(N)})$, and executing the maximum weight search with MWV or MWP-MWV.

There are at most $M^2 L$ vertices in the IDNC graph $\mathcal{G}(\mathcal{V}^{(F)}, \mathcal{E}^{(F)})$. According to condition (c.1), only vertices identified with the same transmission rate are considered for the existence of edges, and there are at most $ML$ vertices identified by the same rate. Thus, using (c.2) to determine the existence of coding edges among the vertices identified by each rate has a complexity of $O(M^2 L^2)$. $M$ possible transmission rates need a total complexity of $O(M^3 L^2)$.

When the MWV algorithm is employed to search the maximum weight clique in $\mathcal{G}(\mathcal{V}^{(F)}, \mathcal{E}^{(F)})$, at most $M^2 L$ vertices are required to update their modified vertex weight according to (19) during each iteration. Since at most $M$ iterations are implemented by MWV (a maximal clique has at most vertices because any two vertices identified by the same receiver cannot belong the same clique), the complexity of the MWV algorithm is $O(M^3 L)$.

If the MWP-MWV algorithm is employed to implement the maximum weight clique search in $\mathcal{G}(\mathcal{V}^{(F)}, \mathcal{E}^{(F)})$ instead of the MWV algorithm, computing the modified vertex with (19) for each vertex in the IDNC graph requires a complexity of $O(M)$, because there are at most $M$ vertices in a MWP according to Lemma 1. Therefore, at most $M^2 L$ vertices in $\mathcal{G}(\mathcal{V}^{(F)}, \mathcal{E}^{(F)})$ requires a total complexity of $O(M^3 L)$.

Therefore, the total complexity of constructing the IDNC graphs $\mathcal{G}(\mathcal{V}^{(F)}, \mathcal{E}^{(F)})$ and searching the maximum weight clique in it by employing MWV or MWP-MWV is $O(M^3 L^2 + M^3 L)$. Similarly, the complexity of creating the IDNC graph $\mathcal{G}(\mathcal{V}^{(N)}, \mathcal{E}^{(N)})$ and implementing the maximum weight searching in it is $O(|\mathcal{M}^{(N)}|^3 L^2 + |\mathcal{M}^{(N)}|^3 L)$ ($|\mathcal{M}^{(N)}| < M$). Thus, the dominated complexity of Algorithm 3 is $O(M^3 L^2)$.

## 6. Simulation results

This section shows the performance of the proposed algorithms in the downlink of a cellular network. The distance dependent large-scale path loss is set to be $128.1 + 37.6 \log_{10}(dis[km])$, and the small-scale fading is assumed to follow independent Rayleigh fading with zero-mean and unit variance. The cell layout is hexagonal with a cell size $\mathcal{R}_c = 500\text{m}$. For simplicity, we assume that if the distance between a receiver and BS is less than $\mathcal{R}_c / 2$, it belongs to the strong receiver group $\mathcal{M}^{(N)}$, otherwise it belongs to the weaker receiver group $\mathcal{M}^{(F)}$. The buffer ratio $\mu$ is defined as the likelihood that a packet is buffered by a receiver. The simulation parameters are summarized in Table I.

TABLE I. Numerical parameters

| Parameters | Value |
|---|---|
| Cellular layout | Hexagonal |
| Cell radius $\mathcal{R}_c$ | 500m |
| Path loss model | $128.1 + 37.6\log10(\text{dis.}[\text{km}])$ |
| Noise power | 174dBm/Hz |
| $R_{\min}$ | 0.4bps/Hz |
| Distribution of users | Uniform |
| Bandwidth | $5\,M\text{Hz}$ |

The performance of the proposed algorithms is compared with the following schemes.

- RLNC: The BS broadcasts the RLNC packets with the maximum transmission power $P_{\max}$. The finite field selected by RLNC is assumed to be large enough, that is, the encoding coefficient vectors among different RLNC packets are linearly independent. The selected transmission rate for each RLNC packet is limited by the minimum achievable capacities among all receivers, i.e., $r^{(RLNC)} = \min_{U_m \in \mathcal{M}} \log_2\left(1 + \frac{P_{\max} \cdot |h_m|^2}{\sigma_m^2}\right)$.

- NOMA-RLNC [18]: Compared with the above RLNC scheme, by employing the NOMA technology at the physical layer, the BS can simultaneously broadcast two RLNC packets $Q^{(F-RLNC)}$ received by all receivers and $Q^{(N-RLNC)}$ received just by the strong receivers by using SIC. The transmission rates for these two RLNC packets are $r^{(F-RLNC)} = \min_{U_m \in \mathcal{M}} \log_2\left(1 + \frac{P^{(F-RLNC)} \cdot |h_m|^2}{P^{(N-RLNC)} \cdot |h_m|^2 + \sigma_m^2}\right)$ and $r^{(N-RLNC)} = \min_{U_m \in \mathcal{M}^{(N)}} \log_2\left(1 + \frac{P^{(N-RLNC)} \cdot |h_m|^2}{\sigma_m^2}\right)$, respectively. Given the total transmission power $P^{(F-RLNC)} + P^{(N-RLNC)} = P_{\max}$ is limited, fractional transmission power allocation (FTPA) with the minimum SNR is adopted.

- IDNC, MWV [1], [3]-[9]: This scheme selects the IDNC packet without considering the physical aspects. The BS broadcasts each IDNC packet with the maximum transmission power $P_{\max}$, and the transmitted rate of the IDNC packet is limited by the minimal achievable capacity of its targeted receivers. The IDNC packet selection problem is reduced to the maximum weight clique problem, and MWV algorithm is adopted to search the maximum weight clique.

- IDNC, MWP-MWV: The difference between this scheme and the above scheme is that it uses the MWP-MWV algorithm instead of the MWV algorithm to search the maximum weight clique.
- R-IDNC, MWV [10]-[14]: This scheme jointly considers transmission rate adaption and IDNC packet selection with the maximum transmission power $P_{\max}$. The optimization problem is formulated to the maximum weight clique problem and the MWV algorithm is developed to solve the maximum weight clique problem.
- R-IDNC, MWP-MWV: Compared with the above R-IDNC scheme employing MWV, this scheme employs the MWP-MWV algorithm to jointly select the IDNC packet and transmission rate instead of the MWV algorithm.

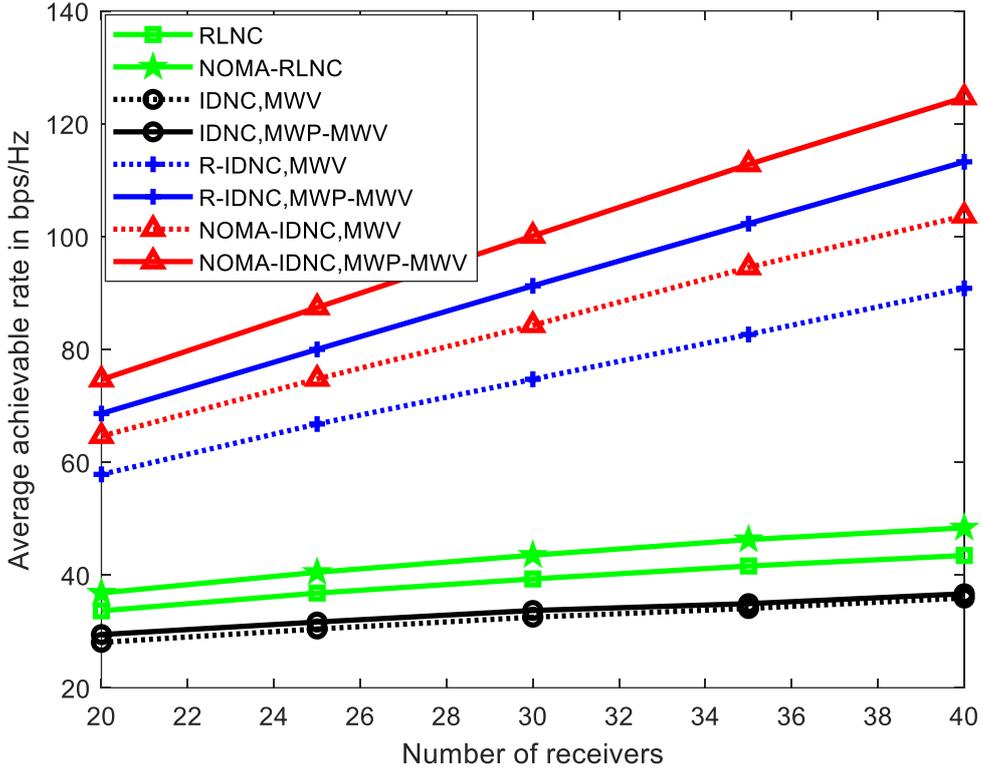

**Fig. 5.** Average throughput vs. the number of receivers $M$.

In Fig. 5, we plot the average throughput versus the number of receivers $M$ with the maximum transmit power $P_{\max} = -42.6 \ dBm/Hz$, the number of packets $L = 20$, and the buffer ratio $\mu = 0.6$. From this figure, we can see that when the number of receivers increases, the system throughput also increases. The reason is that a network coded packet can serve a larger number of receivers as the number receivers increases. Moreover, as the number of receivers increases, NOMA-IDNC (NOMA-IDNC employing MWV and NOMA-IDNC employing MWP-MWV) and R-IDNC (R-IDNC employing MWV and R-IDNC employing MWP-MWV) can achieve more performance gains than RLNC, NOMA-RLNC, and IDNC (IDNC employing MWV and IDNC employing MWP-MWV). This is due to the fact that the transmission rates of RLNC, NOMA-RLNC, and IDNC are limited by the minimum achievable capacity among all receivers. For a larger number of receivers, NOMA-IDNC and R-IDNC have more freedom to make a compromise between the transmit rate adaption and the number of receivers that an IDNC packet serves.

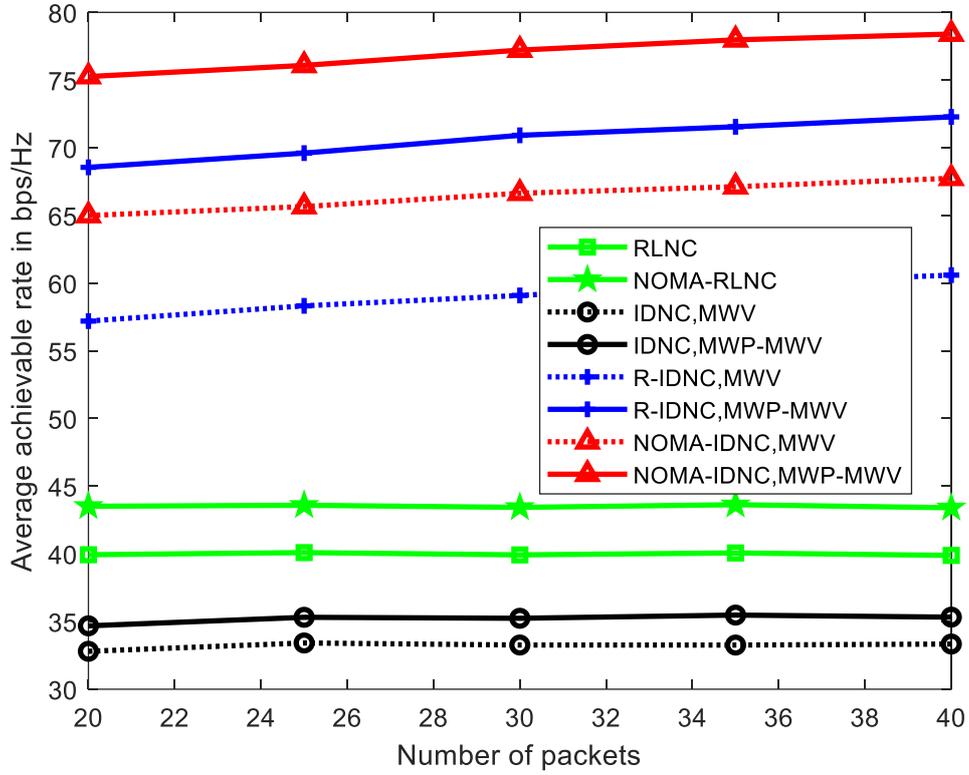

**Fig. 6.** Average throughput vs. the number of packets $L$.

Fig. 6 demonstrate the average throughput versus the number of packets $L$ with the maximum transmit power $P_{\max} = -42.6 \ dBm/Hz$, the number of receivers $M = 20$, and the buffer ratio $\mu = 0.6$. As the number of packets increases, IDNC can increase coding opportunities among the packets. On the other hand, more packets participating in network coding may limits the transmit rate. Similarly, R-IDNC and NOMA-IDNC can adaptively select the transmission rate, achieving a good compromise between coding opportunities selection and transmission rate adaption, and resulting in a slow increase in throughput as the number of data packets increases. Because the considered finite field size of RLNC is large enough, the number of packets does not affect its decoding probability, and therefore it also does not affect the throughput of the RLNC scheme and NOMA-RLNC scheme.

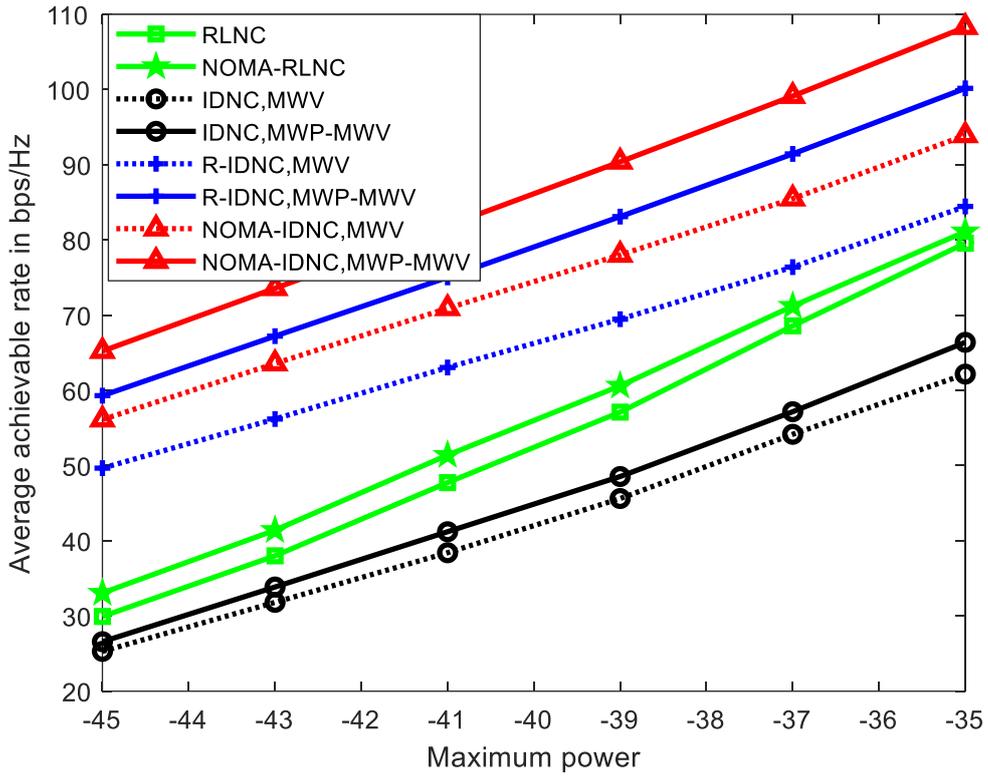

Fig. 7. Average throughput vs maximum power $P_{max}$.

In Fig. 7, we depict the average throughput versus the maximum power $P_{max}$ with the number receivers $M = 20$, the number of packets $L = 20$, and the buffer ratio $\mu = 0.6$. From this figure, we can see that the average throughput of all schemes increases as the maximum power $P_{max}$ increases. This is because the achievable capacities can be increased in this case, which results in a high transmission rate. We can also see from this figure that when $P_{max}$ increases, compare with other schemes, the throughput of RLNC and NOMA-RLNC increases more significantly. This is due to the throughput of RLNC and NOMA-RLNC is limited by the minimum capacity of the receivers.

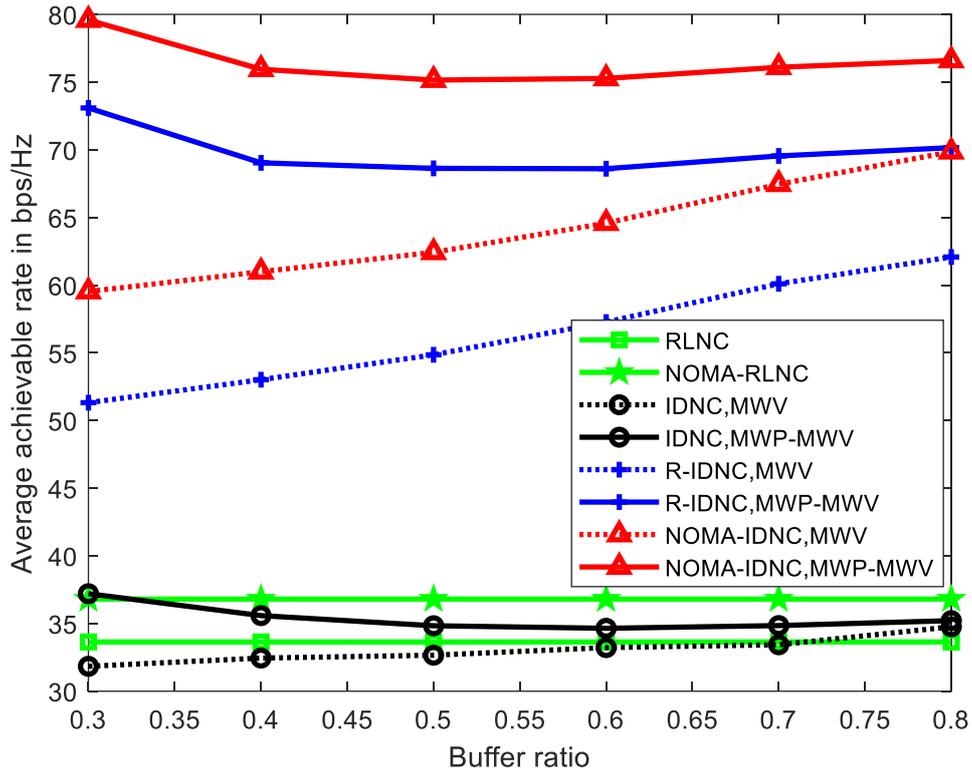

Fig. 8. Average throughput vs the buffer ratio $\mu$.

In Fig. 8, we investigate the impact of the buffer ratio $\mu$ on the average throughput with the maximum transmit power $P_{\max} = -42.6 \ dBm/Hz$, the number receivers $M = 20$, and the number of packets $L = 20$. This figure demonstrates the throughput of the MWV based algorithms (NOMA-IDNC employing MWV, R-IDNC employing MWV, IDNC employing MWV) increases when the buffer ratio $\mu$ increases. When $\mu$ increases, the likelihood of the building an edge between any two vertices in the IDNC graph increases also. Therefore, there are more edges among the adjacent vertices of a vertex, and the modified weight computed with (18) can well indicate the weight of the MWC containing this vertex. The figure also demonstrates that when $\mu$ increases, the throughput of the MWP-MWV based algorithm (NOMA-IDNC employing MWP-MWV, R-IDNC employing MWP-MWV, IDNC employing MWP-MWV) first decreases and then slightly increases. The reason is that when $\mu$ increases, more IDNC coding opportunities can be produced. But on the hand, less packets need to be broadcasted by the BS.

From all of the above figures, we can draw the following observations.

- Our proposed NOMA-IDNC employing MWV (NOMA-IDNC employing MWP-MWV) can achieve better system performance than R-IDNC employing MWV (R-IDNC employing MWP-MWV), and NOMA-RLNC outperforms RLNC. The reason is that NOMA-IDNC (NOMA-IDNC employing MWV and NOMA-IDNC employing MWP-MWV) and NOMA-RLNC can employ NOMA to broadcast two network coded packets during each transmission, so that the strong receivers can receive two network coded packets by employing SIC.
- The MWP-MWV algorithm (NOMA-IDNC employing MWP-MWV, R-IDNC employing MWP-MWV, and IDNC employing MWP-MWV) outperforms the MWV algorithm

(NOMA-IDNC employing MWV, R-IDNC employing MWV, and IDNC employing MWV). As is related in section IV, MWV uses the sum weight of the adjacent vertices of a vertex to indicate weight of the maximum weight clique containing this vertex. When there are less edges among the adjacent vertices (the adjacent vertices cannot form a clique), in (19), the modified vertex weight $\tilde{\omega}(v_{m,l,r})$ cannot reflect the weight of $\omega(\kappa_{\{v_{m,l,r}\}}^{*(F)})$ very well, which results in incorrect selection of the maximum weight clique. In contrast, in (19), MWP-MWV employs $\hat{\omega}(v_{m,l,r})$ to indicate $\omega(\kappa_{\{v_{m,l,r}\}}^{*(F)})$. According to Lemma 1, $\mathbb{L}(v_{l,m,r})$ forms a maximal clique. Therefore, MWP-MWV may better indicates the weight of $\omega(\kappa_{\{v_{m,l,r}\}}^{*(F)})$ than MWV. However, for IDNC, the performance difference between MWP-MWV and MWV is relatively small. In IDNC graph, without considering the physical channel condition, the clique is selected to serve as many receivers as possible, which is equivalent to searching the maximal clique. The throughput of the system is not only related to the number of receivers to be served, but also to the achievable maximal transmission rate. Therefore, whether the maximal clique can be correctly selected does not have a significant impact on the throughput performance for IDNC.
- NOMA-IDNC (NOMA-IDNC employing MWV and NOMA-IDNC employing MWP-MWV) and R-IDNC (R-IDNC employing MWV and R-IDNC employing MWP-MWV) can achieve better performance than RLNC, NOMA-RLNC, and IDNC (IDNC employing MWV and IDNC employing MWP-MWV). The reason is that NOMA-IDNC and R-IDNC can jointly select the transmission rate and IDNC packets to improve the system throughput.
- RLNC can achieve better performance than IDNC. This is because the RLNC packet can serve all receivers, while the IDNC packet may serve a portion of the receivers.

## 7. Conclusions

This paper introduces a cross-layer framework in downlink cellular network to optimize the throughput by jointly considering IDNC at the network layer and NOMA at the physical network layer. The throughput optimization problem is formulated as a joint optimization over IDNC scheduling and power allocation. To make the optimization problem tractable, we decouple it into two subproblems named as IDNC schedule and power allocation. Thus, the two subproblems can be solved separately and iteratively. The IDNC schedule subproblem includes IDNC packets selection and transmission rate adaption. By exploiting the IDNC graph, the IDNC scheduling subproblem can be formulated as the maximum weight clique problem. Two heuristic algorithms named as MWV and MWP-MWV are proposed to solve the maximum weight clique problem. The power allocation subproblem is solved by employing the iterative function evaluation approach. Simulation results verify the effectiveness of our proposed scheme.